# Search for Magnetic Field Expulsion in Optically Driven $K_3C_{60}$

G. De Vecchi[1,*], M. Buzzi[1,*], G. Jotzu[1,*],

S. Fava[1], T. Gebert[1], G. Magnani[2], D. Pontiroli[2], M. Riccò[2], A. Cavalleri[1,3]

[1] Max Planck Institute for the Structure and Dynamics of Matter, 22761 Hamburg, Germany

[2] Dipartimento di Scienze Matematiche, Fisiche e Informatiche, Università degli Studi di Parma, Italy

[3] Department of Physics, Clarendon Laboratory, University of Oxford, Oxford OX1 3PU, United Kingdom

e-mail: michele.buzzi@mpsd.mpg.de, andrea.cavalleri@mpsd.mpg.de

**Photoexcited $K_3C_{60}$ displays several properties reminiscent of equilibrium superconductivity, including transient optical spectra, pressure dependence, and I-V characteristics. However, these observations do not decisively establish non-equilibrium superconductivity, which would be conclusively evidenced by transient Meissner diamagnetism, as shown recently in driven $YBa_2Cu_3O_{6.48}$. Here, we search for transient magnetic field expulsion in $K_3C_{60}$ by measuring Faraday rotation in a magneto-optic material placed in its vicinity. Unlike in the case of homogeneous, insulating $YBa_2Cu_3O_{6.48}$, inhomogeneous, metallic $K_3C_{60}$ powders reduce the size of the effect. With the ~50 nT magnetic field resolution achieved in our experiments, we provide an upper limit for the photo-induced diamagnetic volume susceptibility ($\chi_v > -0.1$). On this basis, we conclude that the photo-induced phase has weaker diamagnetism than superconducting $K_3C_{60}$ at zero temperature. Yet, from recent nonlinear transport measurements in this granular material, we expect a light-induced state similar to the equilibrium superconductor near 0.8 $T_c$, for which $\chi_v > -0.1$. A definitive conclusion on the presence or absence of Meissner diamagnetism cannot be made for $K_3C_{60}$ with the current resolution.**

* These authors contributed equally to this work

Optical manipulation of quantum materials has been extensively deployed in search of out-of-equilibrium functional phases, such as photo-induced ferroelectricity[1,2], magnetism[3–5], charge density wave order[6], topology[7,8], and superconductivity[9–15]. Amongst all these phenomena, the optical enhancement of superconductivity has attracted considerable attention and some controversy.

Excitation of $K_3C_{60}$ with mid-infrared and terahertz optical pulses results in a metastable non-equilibrium phase exhibiting superconducting-like optical conductivity spectra, characterized by a $1/\omega$ divergence of the imaginary conductivity ($\sigma_2$) and a gap in the real part ($\sigma_1$)[12] (Fig. 1a). These features were observed up to base temperatures far exceeding the equilibrium superconducting critical temperature of the sample. They are compatible either with a transient, finite-temperature superconductor with hot quasiparticles or an exotic non-superconducting state with very high mobility[16,17].

Other reports confirm the highly unconventional nature of this light-induced state, which shows superconducting-like pressure scaling of the optical conductivity[13], vanishing two-probe electrical resistance,[14] and nonlinear current-voltage characteristics typical of a finite-temperature granular superconducting state[18,19] (Fig. 1b). These features are rather more indicative of a non-equilibrium light-induced state with superconducting correlations.

As shown for a light-driven cuprate[20], we search for magnetic field expulsion in $K_3C_{60}$ when the sample is photoexcited in a static magnetic field. A non-superconducting photo-induced state in which the mobility is enhanced would not modify a static magnetic field. Conversely, if the material were turned promptly into a superconductor, the magnetic field would be dynamically expelled, in analogy with field-cooled Meissner diamagnetism at equilibrium[21].

## Superconducting-like transport

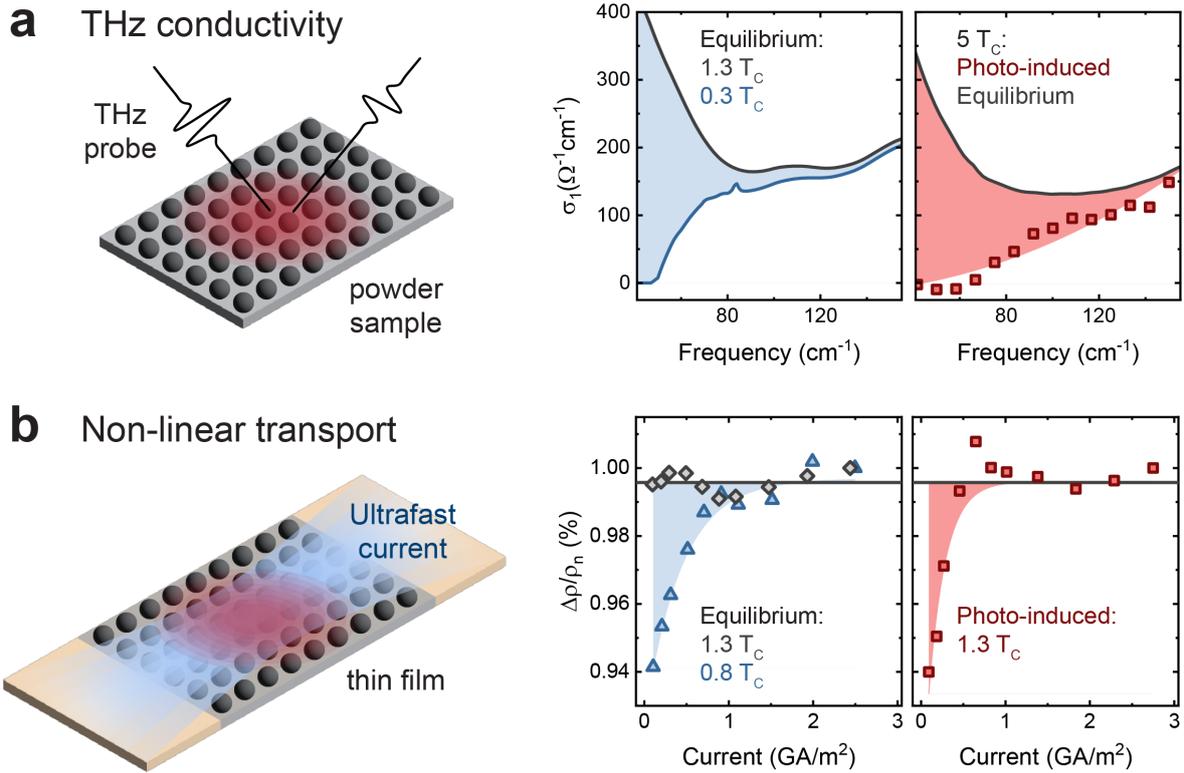

**Figure 1 | Superconducting-like transport properties of photoexcited $K_3C_{60}$.** **(a)** The terahertz conductivity spectra of a $K_3C_{60}$ powder sample are measured using broadband THz pulses (left panel). The real part of the optical conductivity (right panels) is extracted both at equilibrium while cooling across the superconducting transition (dark grey for $T > T_c = 20\,K$, blue for $T < T_c$) and out-of-equilibrium (dark grey line and red symbols refer to spectra measured before photoexcitation and after photoexcitation respectively). A gap opens at equilibrium below $T_c$ and after photoexcitation above $T_c$, indicating dissipation-less transport. Data reproduced from Rowe et al.[31] **(b)** Ultrafast electrical transport is measured in a thin $K_3C_{60}$ film using ultrafast current pulses. The right panel shows the non-linear transport properties of the sample in the equilibrium superconducting state (blue triangles) when increasing the intensity of the ultrafast current pulse. Similar non-linear properties are measured after mid-infrared photoexcitation above $T_c$ (red symbols). These non-linear features are absent at equilibrium above $T_c$ (grey diamonds). Data reproduced from Wang et al.[18]

## Optical Faraday Magnetometry

To track the time- and space-dependent magnetic field surrounding a $K_3C_{60}$ pellet after optical excitation, we employed a time-resolved magneto-optical imaging technique[20,22–24] (Fig. 2a).

This technique is analogous to the one used to measure a transient magnetic field expulsion in driven YBa$_2$Cu$_3$O$_{6.48}$[20]. However, the experiment reported here is more challenging. Indeed, the unperturbed metallic K$_3$C$_{60}$ pellet adjacent to the photoexcited volume is expected to shield ultrafast changes in the local magnetic field. The presence of a time-dependent magnetic flux threading the metallic unperturbed pellet would induce an electromotive force at its boundary and generate eddy currents, counteracting the magnetic flux change through the material. This effect can be described by an equivalent LR low pass circuit (where L is the inductance of the metallic K$_3$C$_{60}$ pellet and R is its resistance). For the geometry relevant to this experiment, the L/R time constant is ~40 ps (see Supplementary Information S11). Hence, regardless of the detector's speed, the response time of magnetic field changes is intrinsically limited by the adjacent metallic sample. Note that this was not the case in the previous experiments[20] in a-c cut YBa$_2$Cu$_3$O$_{6.48}$, for which the low c-axis conductivity strongly reduces a shielding current loop.

A ferrimagnetic Lu$_{3-x}$Bi$_x$Fe$_{5-y}$Ga$_y$O$_{12}$ (hereafter Bi:LIGG) magneto-optic detector[25] was chosen for the present experiments due to its high magnetic field sensitivity. Given the inductive time constant discussed above, the ~100 ps[24,26–28] temporal response of this ferrimagnetic detector is likely appropriate. Importantly, it is smaller than the lifetime of the metastable superconducting-like transport features observed in driven K$_3$C$_{60}$, lasting several nanoseconds[14].

We positioned the detector above the K$_3$C$_{60}$ pellet, as shown in Fig. 2a and Supplementary Information S1 and S2. Bi:LIGG encoded the local magnetic field z-component into a polarization rotation of a linearly polarized probe pulse through the Faraday effect. The spot size of the probe laser limited the achieved spatial resolution. A pellet of ~1 mm diameter and ~30 µm thickness was prepared by pressing K$_3$C$_{60}$ polycrystalline powders in an oxygen-free glovebox to avoid oxidization (Fig. 2b and Supplementary Information

S3). The pellet was sealed between the Bi:LIGG detector and a diamond window, transparent to the mid-infrared excitation pulse (Supplementary Information S4). The entire device is schematically illustrated in (Fig. 2a). A Helmholtz coil pair applied a 0.5 mT magnetic field along the z-axis. A linearly-polarized 800 nm ultrashort probe pulse, focused to ~50 μm FWHM, was directed at near normal incidence onto the ~3 μm thick Bi:LIGG detector. After transmission through the Bi:LIGG, the detector's surface closest to the $K_3C_{60}$ pellet reflected the pulse. Analyzing the probe's polarization yielded a measurement of the z-axis component of the local magnetic field, averaged over the volume traversed by the probe laser pulse in the magneto-optic detector. To enhance the intensity of the reflected probe beam and suppress noise from light scattering off the rough $K_3C_{60}$ pellet surface, a ~150 nm thick layer of aluminum was deposited on the side of the magneto-optic detector closest to the $K_3C_{60}$ pellet (Fig. 2a and Supplementary Information S4).

The externally applied magnetic field $B_{app}$ was sinusoidally modulated in time, synchronously to the laser probe pulse train. Signals acquired with $-B_{app}$ were subtracted from those acquired with $+B_{app}$ (see Supplementary Information S6), distilling the magnetic field contributions to the polarization rotation and suppressing potential non-magnetic contributions such as static birefringence.

**Equilibrium Magnetic Field Exclusion**

In these equilibrium magnetic field expulsion measurements, $B_{app}$ was modulated in time, as the $K_3C_{60}$ sample was held in its superconducting state. Therefore, the resulting magnetic response arose from the exclusion of a time-varying magnetic field rather than the expulsion of a static magnetic field. The generated magnetization is equivalent to that measured in zero-field cooling experiments. This effect is related to the perfect conductivity of a superconductor and is distinct from the Meissner effect (field-cooled magnetization). In general, magnetic field exclusion produces a larger signal than magnetic field expulsion as it reduces the impact of trapped flux[21] (see Supplementary Information S7, S8, and S9).

The colored shadings in Fig. 2a represent finite element simulations of the expected changes in the z-component of the magnetic field surrounding the $K_3C_{60}$ pellet. These simulations show a reduction in the magnetic field above the sample (blue regions) as the magnetic flux is excluded from it. The $K_3C_{60}$ pellet was modeled as a homogeneous medium with a negative diamagnetic susceptibility $\chi \sim -0.5$, consistent with values typically measured in zero-field-cooled $K_3C_{60}$ powders (Supplementary Information S9 and S10).

The panel on the left-hand side of Fig. 2b presents a micrograph of a $K_3C_{60}$ pellet, with the red dashed box indicating the region investigated using optical magnetometry. This measurement was performed by raster scanning the device (Fig. 2a) relative to the laser probe beam. The panel on the right-hand side of Fig. 2b displays a two-dimensional map of the measured z-component of the magnetic field when the $K_3C_{60}$ pellet is cooled to T = 5 K, well into its superconducting state ($T_c \sim 20$ K). In this map, a grey line marks the sample edge for reference. A reduction in the local magnetic field (blue) is observed above the sample, in agreement with the simulations shown in Fig. 2a.

Notably, the field map reveals significant spatial inhomogeneity in the degree of magnetic field exclusion across the pellet. Some regions show a field exclusion as high as ~60% (a value which becomes higher at lower values of the externally applied field, Supplementary Information S9), while others show little to none. This result agrees with previous studies that reported granularity and spatial defects in $K_3C_{60}$[18,19,29].

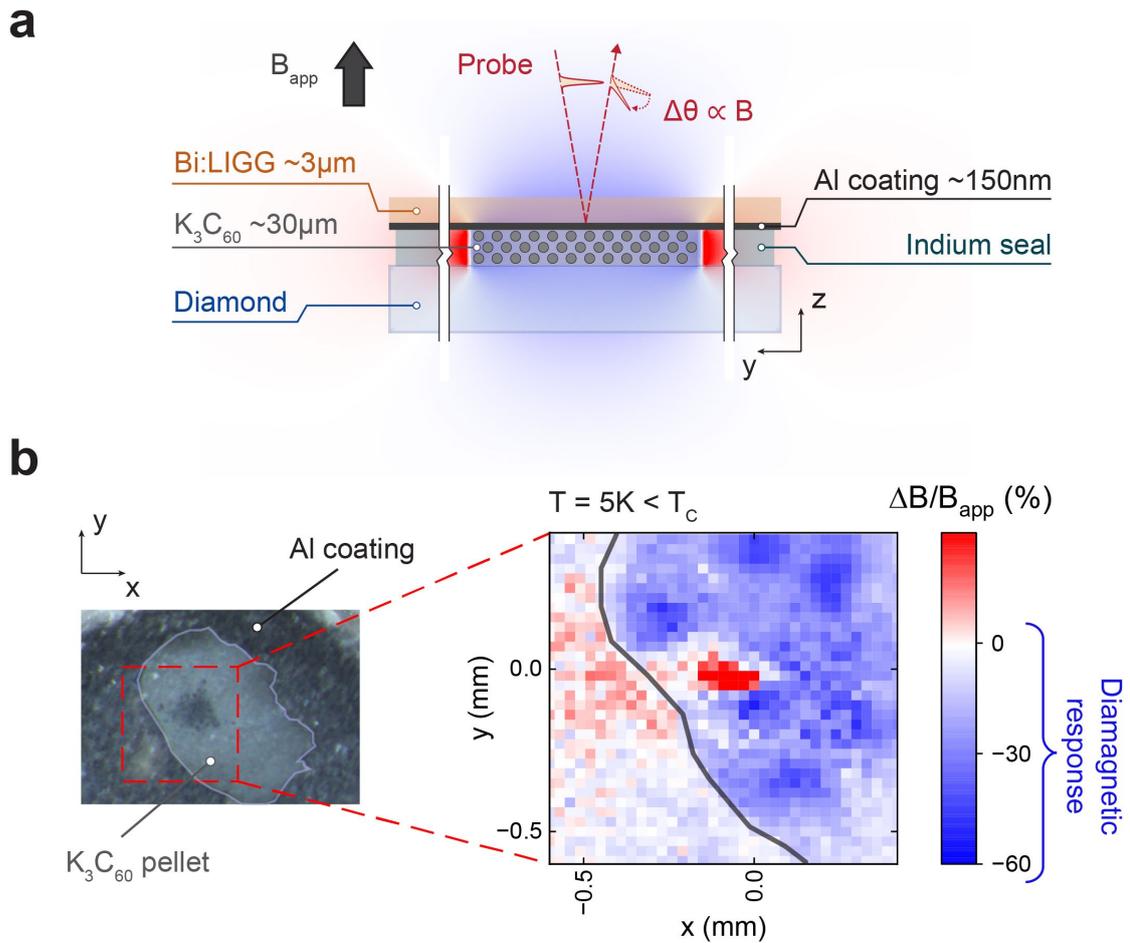

**Figure 2 | Equilibrium Faraday magnetometry. (a)** Sketch of the experimental geometry for Faraday magnetometry. To avoid oxidation, a pellet of pressed $K_3C_{60}$ powder was sealed between a Bi:LIGG magneto-optic crystal and a diamond window. The local magnetic field surrounding the $K_3C_{60}$ pellet was measured by tracking the Faraday polarization rotation of a linearly polarized 800 nm probe pulse, reflected after propagation through the Bi:LIGG crystal. A 0.5 mT magnetic field ($B_{app}$) was applied in the z-direction, and its polarity was periodically cycled to isolate the magnetic contributions to the polarization rotation. A 150 nm thick aluminum coating was deposited on the Bi:LIGG surface to improve the signal-to-noise ratio (Supplementary Information S4). Additionally, the result of a finite element simulation (Supplementary Information S10) highlights the expected magnetic field changes induced by field expulsion in the superconductor (blue indicates reduction and red enhancement of the local magnetic field). **(b)** (left panel) Micrograph of the $K_3C_{60}$ sample on the Al-coated magneto-optic detector. The grey line indicates the edges of the sample, and the dashed red line marks the region where the magnetic field was measured by raster scanning the sample relative to the probe beam. (right panel) Two-dimensional map of the *z*-component of the local magnetic field, measured as a function of the x and y position of the probe at a constant temperature $T = 5K < T_c$. An inhomogeneous reduction of the magnetic field is measured above the sample (blue), indicating the spatially inhomogeneous magnetic field shielding. The black line indicates the outline of the $K_3C_{60}$ sample.

**Search for Magnetic Field Expulsion in Optically Driven $K_3C_{60}$**

The same $K_3C_{60}$ pellet was then excited with a ~1 ps long, 8 µm wavelength mid-infrared pulse (Supplementary Information S5), which is known to induce metastable superconducting-like THz transport properties[14] and other electrical features discussed above. Fig. 3a illustrates the experimental geometry, where the photoexcited region (marked in red) has lateral dimension and thickness equal to the mid-infrared pump beam cross-section and its penetration depth in $K_3C_{60}$, respectively (Supplementary Information S11).

If a diamagnetic response emerges following photoexcitation in $K_3C_{60}$, the resulting spatial profile changes in the magnetic field surrounding the photoexcited region should resemble the finite element simulations displayed in Fig. 3a. Specifically, a reduction of the local magnetic field should be observed above the center of the photoexcited region.

The pump and the probe beam were positioned at a location in the $K_3C_{60}$ pellet for which the best field exclusion was observed in the equilibrium measurements (Fig. 2b). The sample was then photoexcited at a fluence of 40 mJ/cm$^2$, starting from its normal metallic state at T = 50 K > $T_c$ = 20 K.

Fig. 3b displays the measured magnetic dynamics as a function of pump-probe delay for an externally applied field $B_{app}$ = 0.5 mT. Within the current sensitivity of our experimental setup, no statistically significant signal larger than ~50 nT could be detected up to 1 ns time delay.

This result should be evaluated by comparing it with the magnetic field dynamics generated by a disc of the same size as the photoexcited volume acquiring a finite intrinsic diamagnetic susceptibility $\chi_i$ after photoexcitation. As a possible value for $\chi_i$, we take that measured in $K_3C_{60}$ pellets at equilibrium after field-cooling. Without knowledge about the granularity and the density of defects in the photoexcited state and given the significant sample-to-sample variability, we used statistical values extracted from a set of pellets similar to the one considered here to estimate $\chi_i$. We measured the field-cooled intrinsic equilibrium diamagnetic susceptibility of six pellet samples using a commercial magnetometer (Supplementary Information S9). The temperature dependence of $\chi_i$ averaged across all samples is displayed in Fig. 3c (filled diamonds). The shaded bands represent the standard deviation of the sample population (grey) and the range between maximum and minimum values at each temperature (light grey).

We then simulated the expected magnetic field dynamics under two scenarios: (i) a metastable diamagnetic response emerging after photoexcitation and (ii) a diamagnetic response persisting only for the duration of the drive. These finite element simulations, presented in Fig. 3(d,e), were performed by solving time-dependent Maxwell's equations while accounting for the entire experimental geometry. The photoexcited area was modeled as a medium developing a homogenous, time-dependent magnetic susceptibility $\chi_{\text{photo}}(t)$ (Supplementary Information S11).

In the first scenario, $\chi_{\text{photo}}(t)$ is modeled as a negative step function with ~1 ps fall time reaching values equal to the average field-cooled magnetic susceptibility of the $K_3C_{60}$ pellets at 4 K and 16 K (blue and red line in Fig. 3d, top panel, respectively). The shaded areas represent the range between the maximum and minimum values within the sample population at these temperatures.

The lower panel of Fig. 3d shows the corresponding magnetic field changes. Regardless of the final value $\chi_{\text{photo}}(t \to +\infty)$, the traces show a long-lived magnetic field change after an initial reduction with a time constant of ~150 ps. This slower drop is due to the inductive response of the unperturbed metallic $K_3C_{60}$ pellet and the thin aluminum film deposited on the Bi:LIGG detector (Supplementary Information S11). Given this time scale imposed by inductive currents, the response of the magnetization in the Bi:LIGG magneto-optic detector (~150 ps[24]) does not play a role in determining the measured signal dynamics.

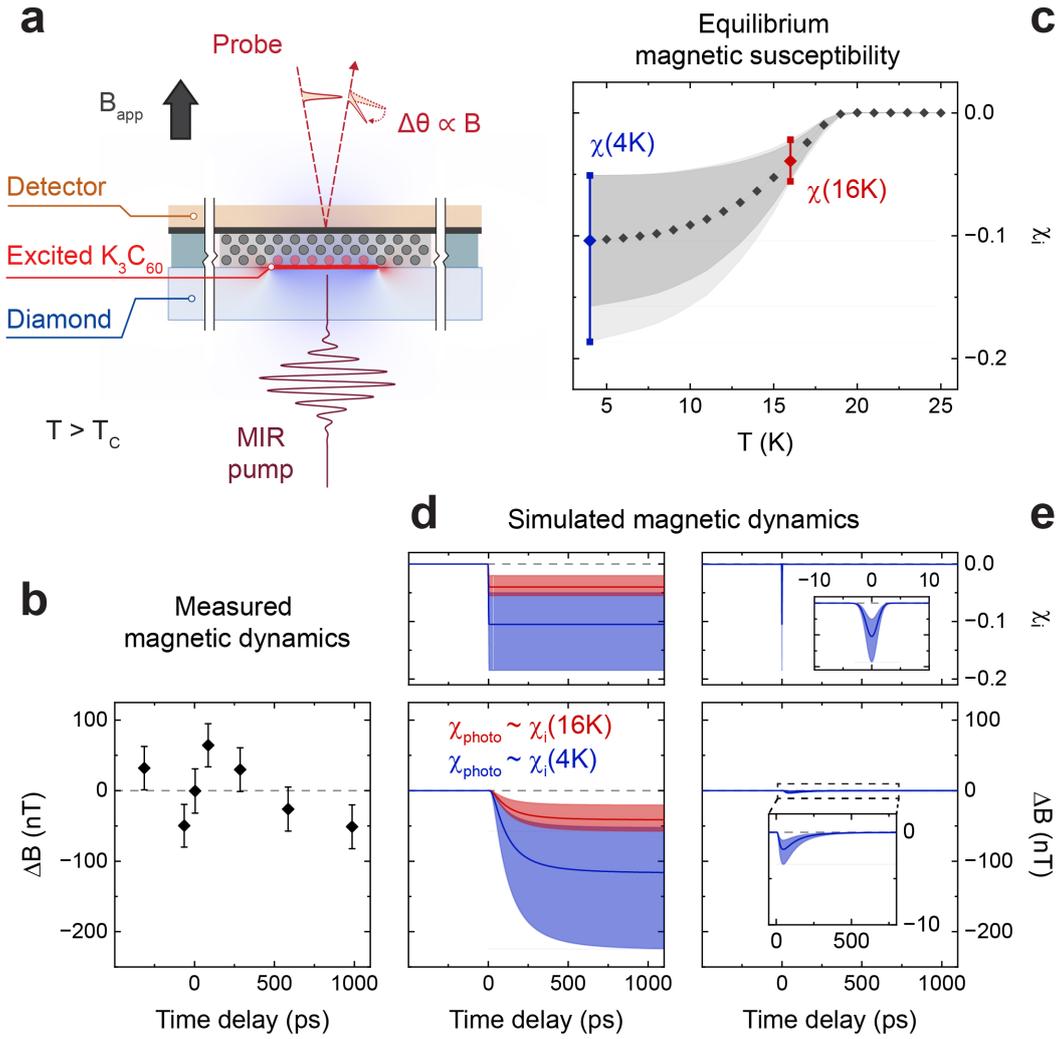

**Figure 3 | Search for magnetic expulsion in photoexcited K$_3$C$_{60}$. (a)** Sketch of the experimental geometry (same as in Fig. 2a). The sample is at a base temperature T = 50 K > $T_c$ and excited with a mid-infrared pump pulse (λ ∼ 8 μm). A possible pump-induced diamagnetic response would manifest as a magnetic field reduction at the detector position (blue shading of the overlayed simulation). The pump-induced changes in the local magnetic field were quantified using Faraday magnetometry in a Bi:LIGG detector as a function of pump-probe time delay. A 0.5 mT magnetic field (B$_{app}$) is applied in the z-direction. Its polarity is periodically cycled to isolate the magnetic contributions to the rotation of the polarization. **(b)** Time dependence of the pump-induced changes in the *z*-component of the local magnetic field ΔB. The error bars denote the standard error of the mean. **(c)** Equilibrium temperature dependence of the diamagnetic susceptibility $\chi_i$ at 1 mT applied field of six pellets of K$_3$C$_{60}$, similar to the one shown in Fig. 2b. The dark grey diamonds represent their average diamagnetic susceptibility, the grey band the interval within one standard deviation from the average, and the light grey band the range between the measured maximum and minimum values (see Supplementary Information S9). **(d)** Step-like quench of the diamagnetic susceptibility (top panel) in the photoexcited region at time zero such that $\chi_{photo}(t \to +\infty) = \chi_i(4\,K)$ (blue) and $\chi_{photo}(t \to +\infty) = \chi_i(16\,K)$ (red). The bottom panels indicate the respective simulated magnetic field dynamics. Shaded bands indicate the range between the maximum and minimum values of $\chi_i$ measured at that temperature. **(e)** Same as in (d) but for $\chi_{photo}(t)$ following a Gaussian pulse peaking at $\chi_{photo}(t = 0) = \chi_i(4\,K)$.

The long-lived magnetic field change is estimated to be ~-115 nT for $\chi_{\text{photo}}(t \to +\infty) = \chi_i(4\,K)$ and ~-40 nT for $\chi_{\text{photo}}(t \to +\infty) = \chi_i(16\,K)$.

In the second scenario, $\chi_{\text{photo}}(t)$ follows a Gaussian pulse peaking at $\chi_{\text{photo}}(t = 0) = \chi_i(4\,K)$ (Fig. 3e, top panel). Due to the inductive response of nearby metallic elements, the maximum amplitude of the magnetic field change is severely reduced to ~-2 nT (Fig. 3e, lower panel), well below our experimental resolution.

Given the results of these simulations and the ultimate sensitivity of our measurements (~30nT, as determined by the standard error of the mean shown in Fig. 3b), we would be able to detect a light-induced magnetic field change only if photoexcited $K_3C_{60}$ entered a metastable state with a diamagnetic susceptibility equal to that measured in the 4 K equilibrium superconducting state. We note that the non-linear current measurements of Fig. 1b, performed under similar excitation conditions, showed that the measured superconducting-like electrical features are equivalent to those seen at equilibrium close to $T_c$ (T=16 K)[18]. According to our time-dependent simulations, the present experimental resolution is not sufficient to detect a diamagnetic susceptibility corresponding to the equilibrium 16 K response.

**Discussion**

The present experiment sets a limit to the maximum magnetic field expulsion that could be observed in driven $K_3C_{60}$ using the best ultrafast magnetometry technique achievable to date. The experimental conditions are unfavorable due to the geometry of the photoexcited state, its distance from the magneto-optic detector, and the inductive response of the unperturbed metallic $K_3C_{60}$. Although we achieved sensitivity to magnetic field changes of a few tens of nanotesla, the upper limit we can set for the magnitude of the diamagnetic susceptibility of the excited region is of order $10^{-1}$. This value is relatively high, even for an equilibrium type-II superconductor, but lower than the value measured in driven YBa$_2$Cu$_3$O$_{6.48}$[20]. For reference, the magnitude of the equilibrium magnetic susceptibility of some K$_3$C$_{60}$ samples we measured is lower than this limit, even at temperatures as low as 0.2 T$_c$ (Fig. 3c).

Furthermore, it is becoming clear that quasi-particles are present in the photo-excited state[16–19,30] and that granularity and defects are an important feature in K$_3$C$_{60}$. Therefore, we expect the photo-induced diamagnetic response to be weaker than that of its low-temperature equilibrium counterpart.

The development of new, more sensitive, time-resolved magnetometry techniques or advances in sample preparation could further improve the sensitivity of this experiment. Additionally, performing this measurement at 10 THz excitation frequency might quantitatively improve the characteristics of the photoexcited state[31].

## Acknowledgments

We acknowledge support from the Deutsche Forschungsgemeinschaft (DFG) by the Cluster of Excellence CUI: Advancing Imaging of Matter (EXC 2056, project ID 390715994). We thank Michael Volkmann, Issam Khayr and Peter Licht for their technical assistance. We are also grateful to Boris Fiedler, Birger Höhling and Toru Matsuyama for their support in the fabrication of the electronic devices used on the measurement setup, to Elena König, Guido Meier, and Eryin Wang for help with sample fabrication and characterization.

# Search for Magnetic Field Expulsion in Optically Driven $K_3C_{60}$

G. De Vecchi[1,*], M. Buzzi[1,*], G. Jotzu[1,*],
S. Fava[1], T. Gebert[1], G. Magnani[2], D. Pontiroli[2], M. Riccò[2], A. Cavalleri[1,3]

[1] Max Planck Institute for the Structure and Dynamics of Matter, 22761 Hamburg, Germany
[2] Dipartimento di Scienze Matematiche, Fisiche e Informatiche, Università degli Studi di Parma, Italy
[3] Department of Physics, Clarendon Laboratory, University of Oxford, Oxford OX1 3PU, United Kingdom
e-mail: michele.buzzi@mpsd.mpg.de, andrea.cavalleri@mpsd.mpg.de

## Supplementary Information

* These authors contributed equally to this work

## S1. Ultrafast Magnetometry Technique

Ultrafast optical magnetometry relies on the Faraday effect, which directly relates the magnetic field applied to a material to the polarization rotation of a linearly polarized beam traversing the medium. This relation is usually reported as:

$$\theta = V \cdot \int_0^L B(z) dz \qquad \text{(Eq. S1)}$$

where θ represents the rotation of the polarization of the input beam, B(z) is the magnitude of the magnetic field along the light propagation direction inside the medium, and L is the thickness of the medium. The proportionality constant V is known as the Verdet constant, a material-dependent constant that also depends on other parameters, such as the wavelength of the incoming polarized light. This expression highlights the fact that the magnetic field is averaged across the thickness of the material traversed. Therefore, if B varies strongly over distances of order ~L, its spatial dependence is blurred in the accumulated polarization rotation.

In the past, the Faraday effect in ferromagnetic crystals and thin films has been used to image the magnetic properties of superconductors at equilibrium[1,2]. These types of detectors (such as Bi:$R_3Fe_5O_{12}$, EuS, and EuSe) offer very high sensitivity (V~$10^5$ rad·T$^{-1}$·m$^{-1}$) but have limited time resolution, down to 100 ps at best[3], due to the presence of low-lying magnetic excitations (e.g., ferromagnetic resonance) at sub-THz frequencies. On the other hand, diamagnetic II-VI and III-V semiconductors such as ZnSe, ZnTe, and GaP have a magneto-optic response featuring Verdet constants that are two to three orders of magnitude smaller than those observed in ferromagnetic materials but have the advantage of not being magnetically ordered and offer significantly better time resolution[3–6].

Given the high conductance of the $K_3C_{60}$ pellet measured in this work, picosecond time resolution was unnecessary (see Supplementary Information S11), and we opted for an improved magnetic field sensitivity. The measurements shown throughout the manuscript were performed using a $Lu_2Bi_1Fe_4Ga_1O_{12}$ (Bi:LIGG, see Supplementary Information S2) magneto-optic crystal. Its sensitivity to magnetic fields was calibrated before each measurement in the exact experimental conditions used for the measurement. We measured the amount of angular polarization rotation in a position of Bi:LIGG far from the sample (see Fig. 2 of the Main Text), where the magnetic field was equal to the uniform externally applied magnetic field. In this way, we could extract a calibration factor specific to our experimental conditions, which allowed us to convert the measured polarization rotation into the value of the local magnetic field in the crystal.

Finally, to have an accurate calibration of the actual value of the magnetic field, the current-to-magnetic field constant of the Helmholtz coil pair was independently calibrated in situ using a Lakeshore 425 gaussmeter. These calibration measurements yielded a Verdet constant of ~$0.5 \cdot 10^5$ rad·T$^{-1}$·m$^{-1}$ for Bi:LIGG (Supplementary Information S2).

## S2. Detector Characterization

The ~3μm thick Bi-substituted lutetium iron gallium garnet ($Lu_2Bi_1Fe_4Ga_1O_{12}$, Bi:LIGG for short in the following) magneto-optic detector was obtained through a commercial supplier and grown by Liquid-Phase Epitaxy on a (100)-oriented, 500 μm thick $Gd_3Ga_5O_{12}$ crystal (GGG). The composition of the Bi:LIGG sample was measured using energy-dispersive X-ray spectroscopy (EDX), yielding the following percentage composition of elements: O 58.98%, Fe 19.43%, Ga 5.99%, Lu 10.28%, Bi 5.32%. This result is in good agreement with the chemical formula reported above.

Measurements reported in another manuscript published by some of the authors[3] reveal the presence of a low-frequency magnon (centered at a frequency of ~ 6 GHz) in the same Bi:LIGG compound used here as a magneto-optic detector. The time response of the material's magnetization, and consequently the response of the Faraday polarization rotation, is limited to the inverse of this excitation frequency ~ 160 ps. This time resolution does not constitute an issue for the measurement we want to perform here due to the intrinsic low-pass filtering effect due to the magnetic shielding from the metallic $K_3C_{60}$ pressed pellet (Supplementary Information S11).

## S3. Sample Growth, Characterization and Preparation

The $K_3C_{60}$ powder used for the pellets discussed in this work was prepared and characterized as reported previously[7–10]. Stoichiometric amounts of ground $C_{60}$ powder and potassium were placed in a sealed Pyrex vial, which was evacuated to $10^{-6}$ mbar pressure. Whilst keeping the $C_{60}$ powder and solid potassium separated, the vial was kept at 523 K for 72 h and then at 623 K for 28 h so that the $C_{60}$ powder was exposed to pure potassium vapor. The vial was then opened inside an Ar glovebox (<0.1 ppm $O_2$ and $H_2O$), where the powder was reground and pelletized before annealing at 623K for 5 days. X-ray diffraction measurements were then carried out on the resulting $K_3C_{60}$ powder, confirming that it was phase pure, with an average grain size between 100 and 400 nm. The quality of the sample was then checked by producing a pellet (see discussion below) and measuring its magnetic susceptibility (see Supplementary Information S9). These measurements revealed a significant variability across the different pellets measured, even if the pellet preparation process was reproduced consistently for each measurement.

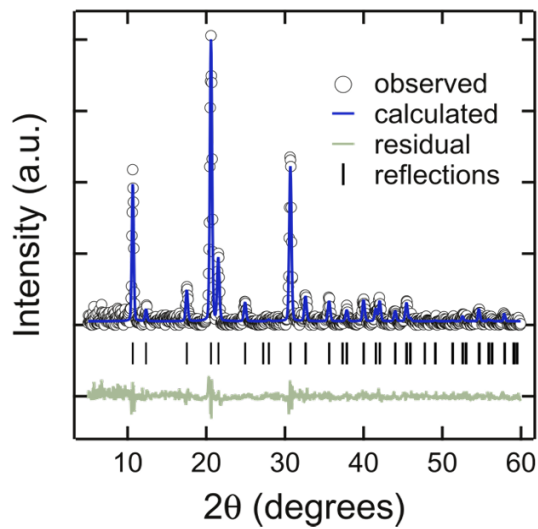

**Figure S3.1** | X-ray diffraction data and single f.c.c. phase Rietveld refinement for the $K_3C_{60}$ powder used in this work.

This $K_3C_{60}$ powder was then used to prepare the pellet measured in the Main Text. The powder was pressed between two press dies in a polypropylene mask to control the resulting pellet's thickness and shape (Fig. S3.2).

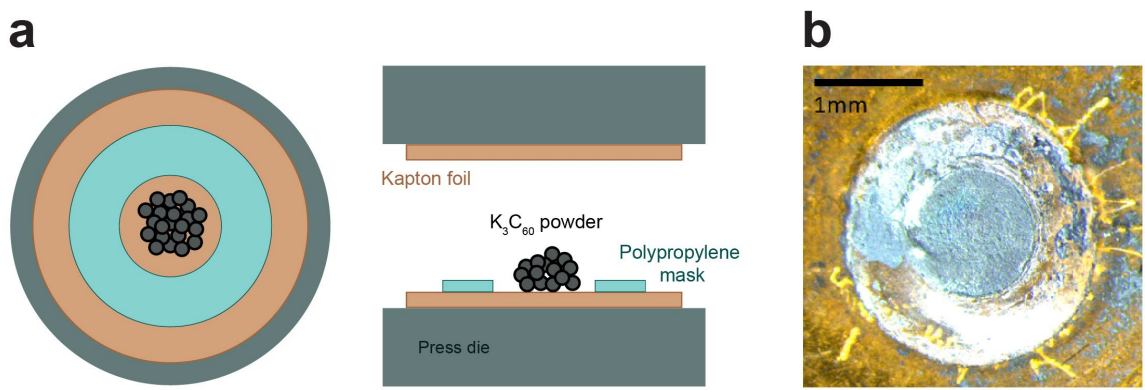

**Figure S3.2** | Process to produce a $K_3C_{60}$ pellet of known thickness. **(a)** Sketch representing the preparation phase. Some $K_3C_{60}$ powder is placed in the center of a polypropylene ring 30 μm thick on top of a Kapton foil disk. This disk prevents the $K_3C_{60}$ powder from sticking to the metallic surface of the press die. The top view is on the left panel, while the side view is on the right panel. **(b)** Micrograph showing a 1 mm diameter pellet after pressing the powder in the polypropylene mask.

Additionally, the surface of each press die was covered with Kapton foil, which prevented the powder sample from sticking to the metallic surface after pressing. The whole sample preparation procedure was performed inside an Ar glovebox (<0.1 ppm $O_2$ and $H_2O$) to avoid oxidization of the sample. All the tools and surfaces coming in contact with the sample were cleaned with acetone in an ultrasonic bath, rinsed with isopropanol, and baked for some minutes on a hot plate before being carried inside the glovebox to avoid alteration of the $K_3C_{60}$ properties. An example of a pellet resulting from this process is shown in Fig. S3.2b. The sample shown in Fig. 2 of the Main Text has been prepared with the same procedure. Its irregular shape is due to the edges breaking when removing the pellet from the polypropylene mask.

## S4. Sample Holder and Experimental Geometry

Fig. S4 shows the sample holder (a) and the geometry of the experiment (b). A good thermal contact of the sample was ensured by pressing the $K_3C_{60}$ pellet between the magneto-optic detector and the diamond window. Both the window and the detector were then glued to a sapphire plate using sapphire-based epoxy glue, which had good thermal conductivity and poor electrical transport properties. Finally, the sapphire plates were thermally contacted with a cold finger to cool down the sample holder. The Bi:LIGG detector was coated with a 150 nm thick aluminum layer to maximize the intensity of the light propagating through the magneto-optic crystal and and to suppress diffuse scattering from the rough $K_3C_{60}$ surface that would have cause additional noise. Part of the detector was not coated with aluminum (Fig. S4c).We used the sharp edges of the Al-coating to perform knife-edge scans and precisely identify the position and size of the probe beam in situ. Furthermore, by letting the intense mid-infrared pump propagate with the probe beam in the uncoated region, we could find the time overlap between the pump and the probe. Conversely, the sample and the metallic film completely prevented this cross-talk between the pump and probe by absorbing or reflecting the pump radiation. The sample holder was designed to minimize the presence of electrically conductive loops, in which eddy currents would build up to shield rapid changes of the magnetic field (Supplementary Information S8). The only highly conductive components close to the photo-excited region are the puck of pressed $K_3C_{60}$ powder, which is metallic with a conductivity of $\sim 10^5$ S/m, the aluminum coating deposited on the Bi:LIGG detector, and the sealing indium ring shown in Fig. S4c. We discuss the effect of the metallic $K_3C_{60}$ puck and the aluminum film in Supplementary Information S11. The presence of the indium ring can be neglected. Due to its the poor mutual inductance with the much smaller photo-excited region, this indium ring shields time dependent changes of magnetic field generated by the photo-excited region by less than one part per million.

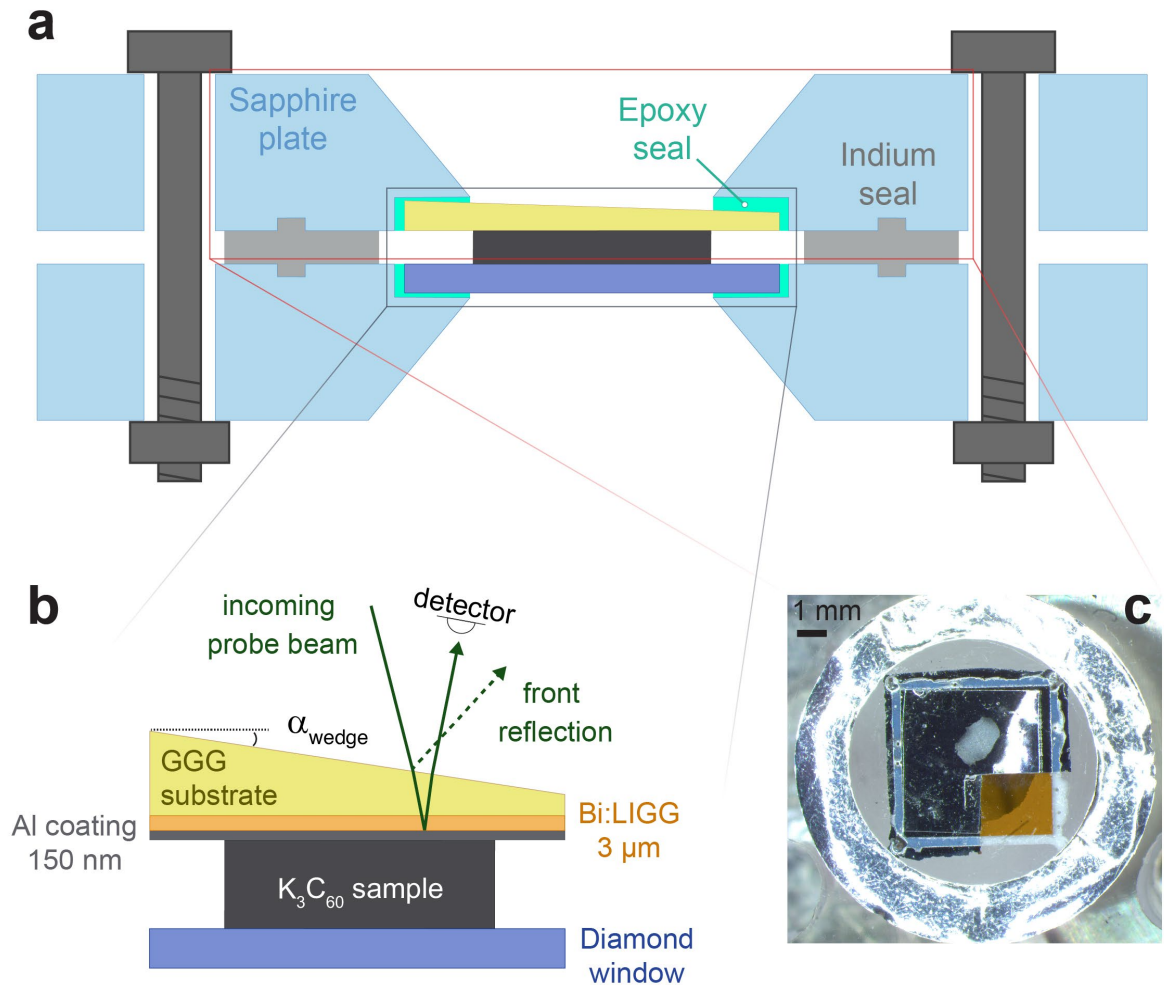

**Figure S4 | (a)** Schematic representation of a section of the sample holder. The substrate of the magneto-optic detector (yellow) and the diamond window (dark blue) were each glued to a sapphire plate using sapphire-based epoxy glue. The $K_3C_{60}$ sample was placed between the magneto-optic detector and diamond window, touching them to ensure thermal contact. A ring of Indium (see (c)) was pressed between the two sapphire plates to seal the sample. The preparation of this sample holder was performed inside an Ar glovebox to avoid sample contamination. **(b)** Side view of the experimental geometry. The GGG substrate of the Bi:LIGG detector was wedged at an angle $\alpha_{wedge} \sim 1.5°$ (Supplementary Information S2) to separate the front and back reflections from each other spatially and only collect the latter with our photo-detector. **(c)** The micrograph shows the front view of the upper sapphire plate with the magneto-optic detector glued to it and coated with aluminum. The indium ring with an internal diameter of 8 mm used to seal the sample is also visible. The sample in the center is the same as in Fig. 2 of the Main Text.

## S5. Experimental Setup and Data Acquisition

The measurements were performed using the experimental setup sketched in Fig. S5. Ultrashort (100 fs) 800 nm center wavelength laser pulses at 900 Hz repetition rate were generated using a pair of commercial Ti:Al$_2$O$_3$ amplifiers seeded by the same oscillator. One amplifier produced ~2 mJ pulses and was used for probing Faraday rotation in the Bi:LIGG detector. The second amplifier produced ~5 mJ pulses and was used to pump a home-built three-stage OPA that generated ~1.5 mJ total energy signal and idler pulses. These pulses were mixed in a 0.4 mm thick GaSe crystal to obtain ~100 fs long, ~40 µJ energy pulses centered at ~40 THz, close to the frequency used in previous works to excite K$_3$C$_{60}$[7–9]. These pulses were then chirped using a 10 mm CaF$_2$ rod to a duration of ~1 ps to deposit more energy on the sample without damaging it and induce the metastable superconducting state revealed in reference [9].

The polarization of the probe beam was set using a nanoparticle high-extinction ratio linear polarizer to minimize the noise sources in the measurement. As non-normal incidence reflections introduce a phase delay between *s* and *p* polarization, incidence angle fluctuations can give rise to polarization noise. Only reflections close to normal incidence were used in the setup to minimize the effect of angular fluctuations, and a commercial system using active feedback was used to stabilize the laser beam pointing. After traversing and being reflected from the second surface of the Faraday detector, the polarization state of light was analyzed using a half-waveplate, Wollaston prism, and balanced photodiode setup that allowed us to quantify the Faraday effect in the magneto-optic detection crystal. The K$_3$C$_{60}$ sample was embedded in the detector assembly (Fig. S4) and mounted on the cold finger of a liquid helium cryostat to allow for temperature control. The cryostat was directly placed in a high vacuum chamber. A pair of coils in a Helmholtz configuration generated a magnetic field at the sample position whose polarity could be reversed at a frequency of 450 Hz. Switching the polarity of the magnetic field helped reject possible spurious contributions (Supplementary Information S6) arising from unwanted sources of bi-refringence (e.g., residual strain in the detector crystals or vacuum windows). The highest achievable magnetic field was limited by heat dissipation and was ~3 mT. The sample position was changed using computer-controlled linear translation stages that made it possible to reproducibly move the cryostat and the sample inside the vacuum chamber with ~10 µm repeatability.

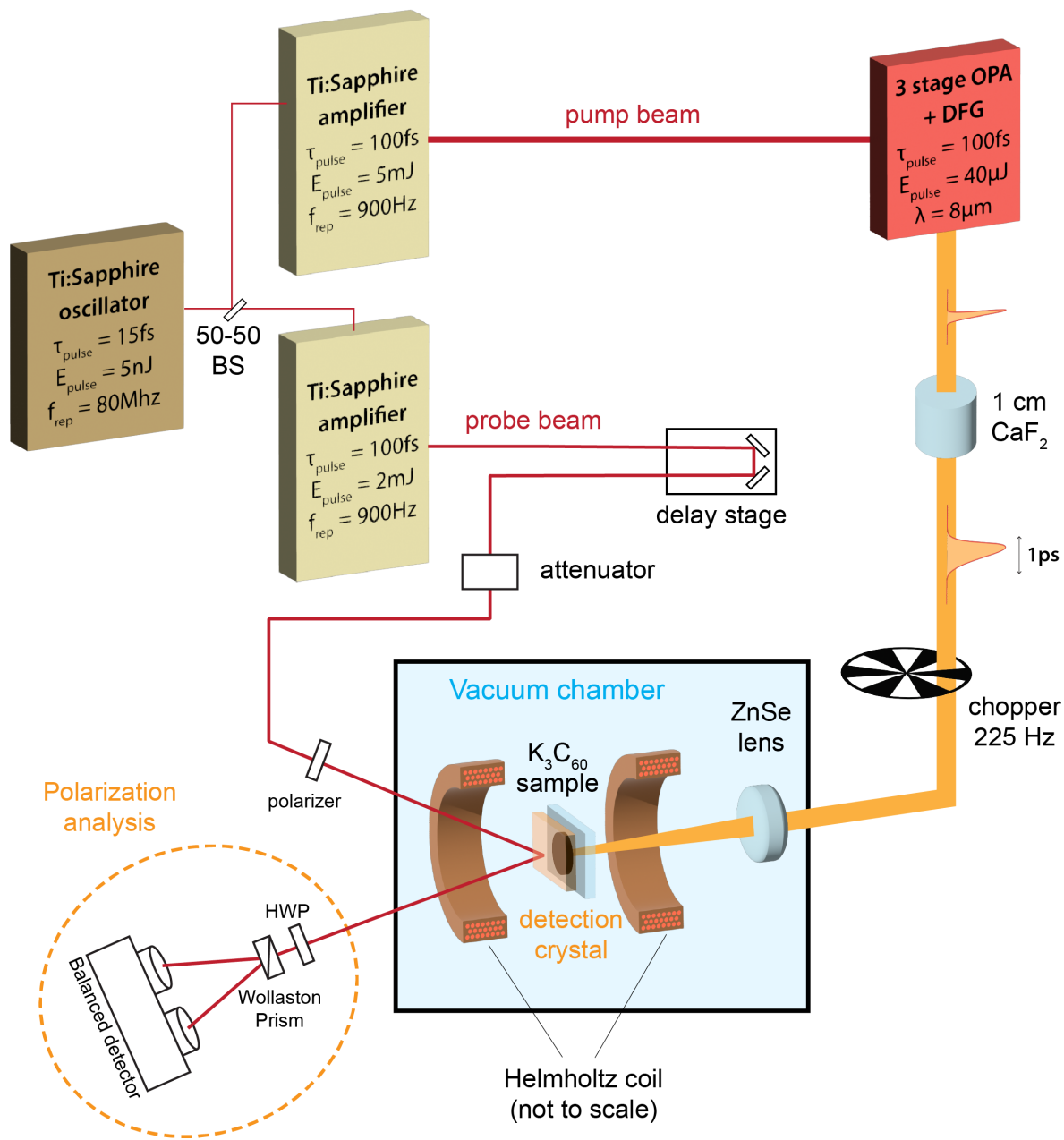

**Figure S5** | Experimental setup.

The electrical pulses from the balanced photodetector were digitized using a commercial 8-channel 40MS/s data acquisition card triggered at the lowest frequency used in the experiment to obtain differential magnetic field measurements. These pulses, acquired in the time domain, were then integrated by applying a boxcar function to yield the signals from the sum and difference channels of the balanced photodetector for each probe laser pulse (see Fig. S6.1-2). Since acquiring a complete pulse sequence required the acquisition of many cycles of the applied magnetic field, the sample clock signal of the data acquisition card was derived using direct digital synthesis from the oscillator repetition rate. In this way, drifts in the cavity length and repetition rates of the system did not affect the relative timing of the boxcar functions compared to the arrival time of the electrical signal from the photodiodes.

## S6. Data Reduction and Analysis

As mentioned in the previous section, the polarity of the magnetic field was cycled periodically, and probe pulses with pump and without pump were acquired to yield double differential pump-probe measurements and isolate only magnetic contributions to the polarization rotation, which inverts with the applied magnetic field. In other words, because the equilibrium and pump-induced magnetic field changes measured with applied field -$H_{app}$ were subtracted from those acquired with applied field +$H_{app}$, the sensitivity of the signal to strain in the magneto-optic detector, polarization noise in the setup or to possible non-magnetic pump-induced signals was strongly reduced.

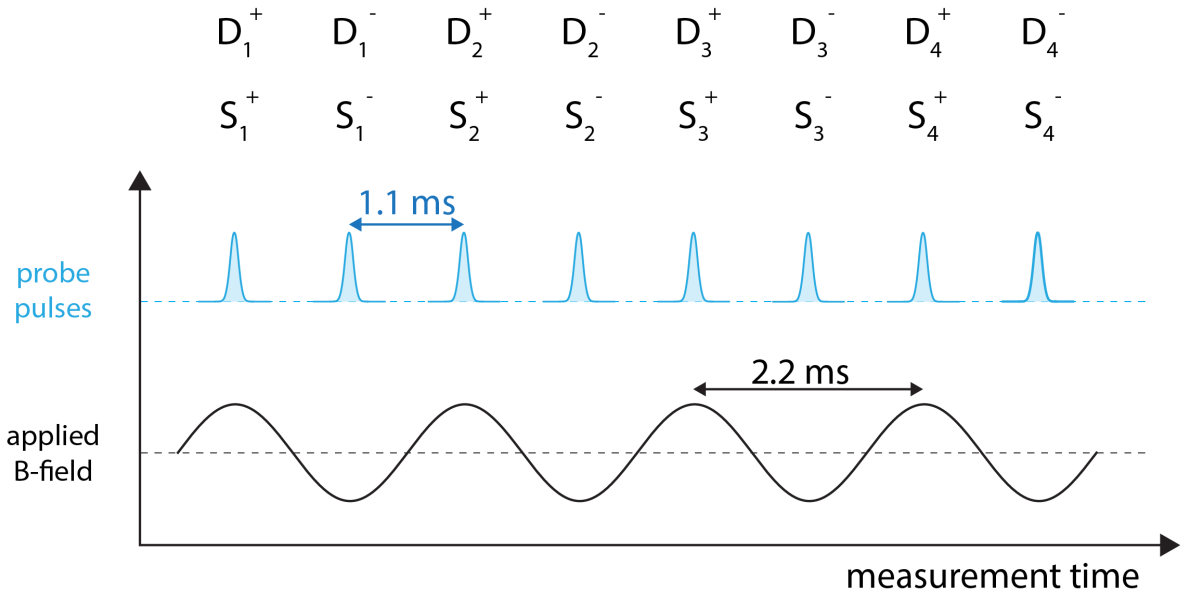

**Figure S6.1** | Timing diagram of the acquisition scheme used for the equilibrium measurements shown in Fig. 2b of the Main Text and Fig. S9.3. Signals acquired by the difference and sum channels of the photodetector are marked with D and S, respectively.

The magnetic field polarity was cycled following a sinewave at 450 Hz frequency. We chose the highest possible switching frequency to reject low-frequency 1/f noise efficiently. A timing diagram of the acquisition scheme for equilibrium measurements is shown in Fig. S6.1. As mentioned in Supplementary Information S5, we triggered the data acquisition card at a sub-harmonic frequency of the probe repetition rate corresponding to traces comprising 88 pulses (a trace of 8 pulses is shown in Fig. S6.1 as an example). In the end, only 80 of these 88 pulses were acquired due to a fixed dead time for data processing.

We label signals from the difference and sum channels of the photodetector as $D_i^{\pm}$ and $S_i^{\pm}$, respectively, to indicate whether they were acquired with positive (+) or negative (-) polarities of the applied magnetic field. The subscript *i* runs over the 80 acquired pulses in the acquisition trace. The equilibrium polarization rotation within a trace is calculated in the following way:

$$\Delta\vartheta_{eq} = \frac{\sum_i^{80} D_i^+}{\sum_i^{80} S_i^+} - \frac{\sum_i^{80} D_i^-}{\sum_i^{80} S_i^-}. \qquad (Eq.\ S6.1)$$

This quantity yielded the amplitude of the magnetic field after calibration of the Faraday effect in the Bi:LIGG detector (see Supplementary Information S1). To limit computational dead times between acquisitions, we normalize after averaging as indicated in the formula above.

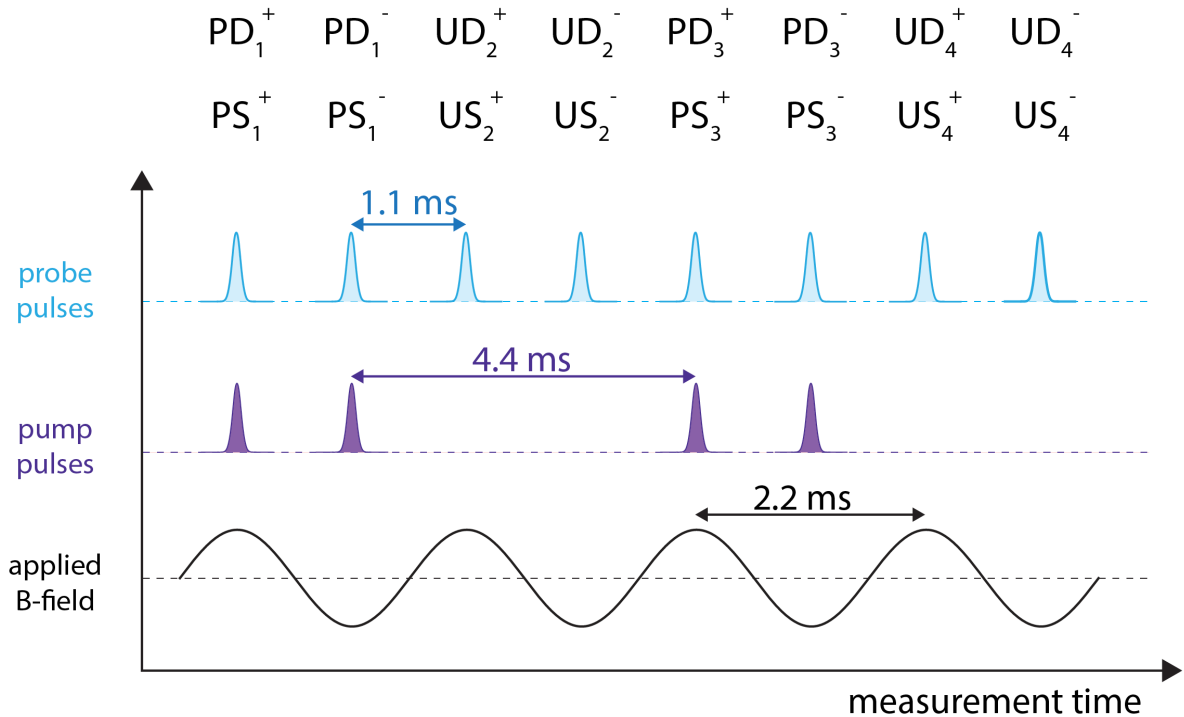

**Figure S6.2** | Timing diagram of the acquisition scheme used for the pump-probe measurement shown in Fig. 3b of the Main Text. Signals acquired with and without the pump are labeled with P and U, respectively. Similarly, the difference and sum channels of the photodetector are marked with D and S, respectively.

Similarly, in pump-probe experiments, the pump was mechanically modulated at 225 Hz (Fig. S6.2). In this case, we label signals acquired from the difference and sum channels of the photodetector as $PD_i^{\pm}, PS_i^{\pm}$ when the pump is on ("pumped") and $UD_i^{\pm}, US_i^{\pm}$ when the pump is off ("unpumped"). Again, the labels "+" and "-" indicate the polarities of the externally applied magnetic field.

The pump-probe polarization rotation within an acquisition trace is calculated in the following way:

$$\Delta\vartheta_{p-p} = \frac{\sum_i^{80} PD_i^+}{\sum_i^{80} PS_i^+} - \frac{\sum_i^{80} PD_i^-}{\sum_i^{80} PS_i^-} - \frac{\sum_i^{80} UD_i^+}{\sum_i^{80} US_i^+} + \frac{\sum_i^{80} UD_i^-}{\sum_i^{80} US_i^-}. \quad \text{(Eq. S6.2)}$$

Using the same calibration factor for the Faraday effect used for the equilibrium measurements, this quantity yielded the amplitude of the pump-induced magnetic polarization rotation.

The sequential acquisition of multiple pulses (88) allowed us to minimize the impact of lost pulses during processing time while keeping the frequency of acquisition as high as possible to reject the effect of low-frequency noise when averaging the difference and sum channels within a trace before taking the ratio of the two quantities. The phase of the sinusoidal magnetic field applied was periodically alternated between 0 and π relative to the start of the acquired laser pulse train to cancel out residual drifts due to possible asymmetries in the applied magnetic field.

## S7. Field Cooled vs. Zero-Field Cooled Magnetization

The magnetic field measured in our experiment originates from changes in the magnetization of $K_3C_{60}$, which can occur through two mechanisms: magnetic field expulsion, the response of a superconductor to static magnetic fields, or magnetic field exclusion, its response to time-varying magnetic fields. These scenarios correspond to the well-known Field-Cooled (FC) and Zero-Field-Cooled (ZFC) measurement protocols.

Both phenomena would result in perfect magnetic shielding below the critical field in a defect-free superconductor. However, as schematically shown in Fig. S7, the presence of defects (depicted as a hole in the superconducting disc) alters this behavior. In the field-cooled scenario, the Meissner effect expels the magnetic field only from the bulk of the superconductor, leaving some of it trapped in the hole and leading to reduced magnetiza-tion[11,12]. In contrast, zero-field-cooled measurements exhibit complete magnetic field exclusion due to the sample's perfect conductivity, as long as a connected loop of supercurrents, surrounding the bulk of the sample, can be formed. To quantify this difference for $K_3C_{60}$, we performed ZFC and FC SQuID magnetometry measurements. The results are reported and discussed in Supplementary Information S9 and reveal the impact of defects on the superconductor's diamagnetic response.

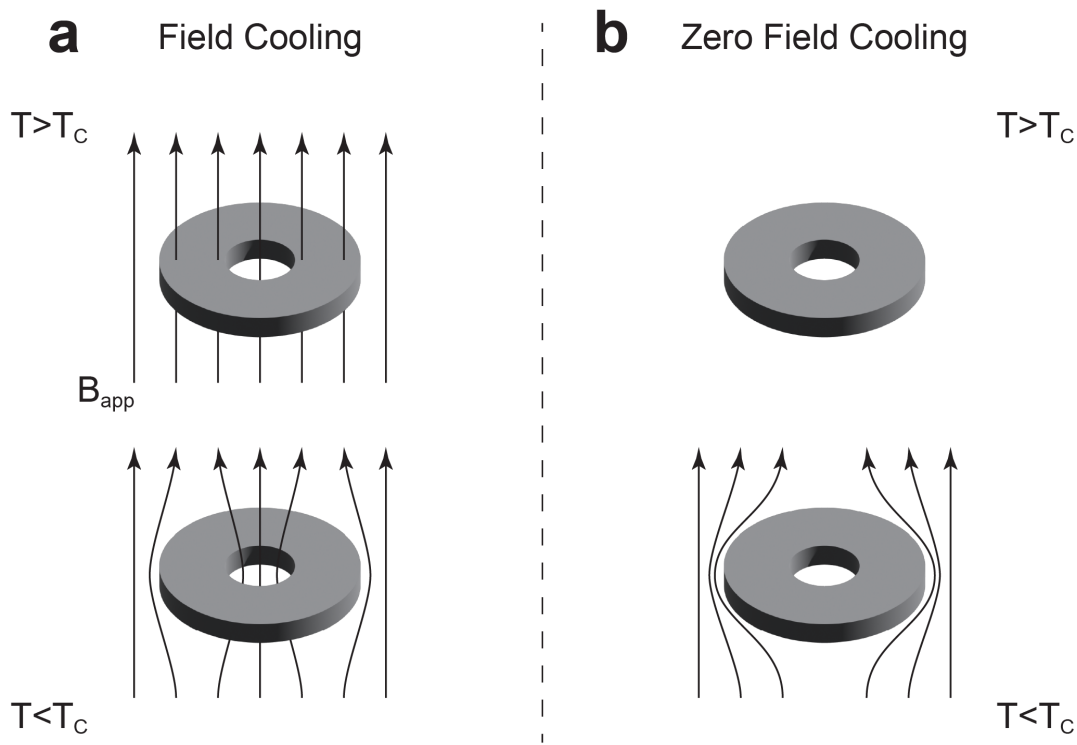

**Figure S7** | The sketch shows how an applied magnetic field is modified by a superconducting sample with a hole, representing a non-superconducting defect. **(a)** When the sample is cooled in a magnetic field, the field is expelled from the sample and trapped into the hole, reducing the net magnetization of the sample compared to a defect-free sample. **(b)** When the sample is first cooled below $T_c$, and then a magnetic field is applied, the field is excluded from the whole sample volume, no matter the presence of defects, as long as a closed loop of supercurrents, surrounding the bulk of the sample, can be formed.

## S8. Shielding of Time-varying Magnetic Fields

This section elaborates on how systems with high electrical conductivity shield time-varying magnetic fields, as the Faraday-Lenz law describes, in the experimental configurations relevant to this manuscript. This concept applies to superconductors below $T_c$ and high-conductivity metals.

### In Superconductors

Magnetic field exclusion occurs when a superconductor is cooled below $T_c$, and a time-varying magnetic field is applied. This phenomenon is due to the perfect conductivity of the $YBa_2Cu_3O_7$ sample below $T_c$. Magnetically induced persistent shielding currents keep the magnetic flux constant in the superconductor, as the Faraday-Lenz law prescribes. The magnetic field is entirely excluded from the sample volume, even if defects exist, unlike what happens for the Meissner effect (Supplementary Information S7). This shielding response is observed in zero-field cooled experiments[11,12].

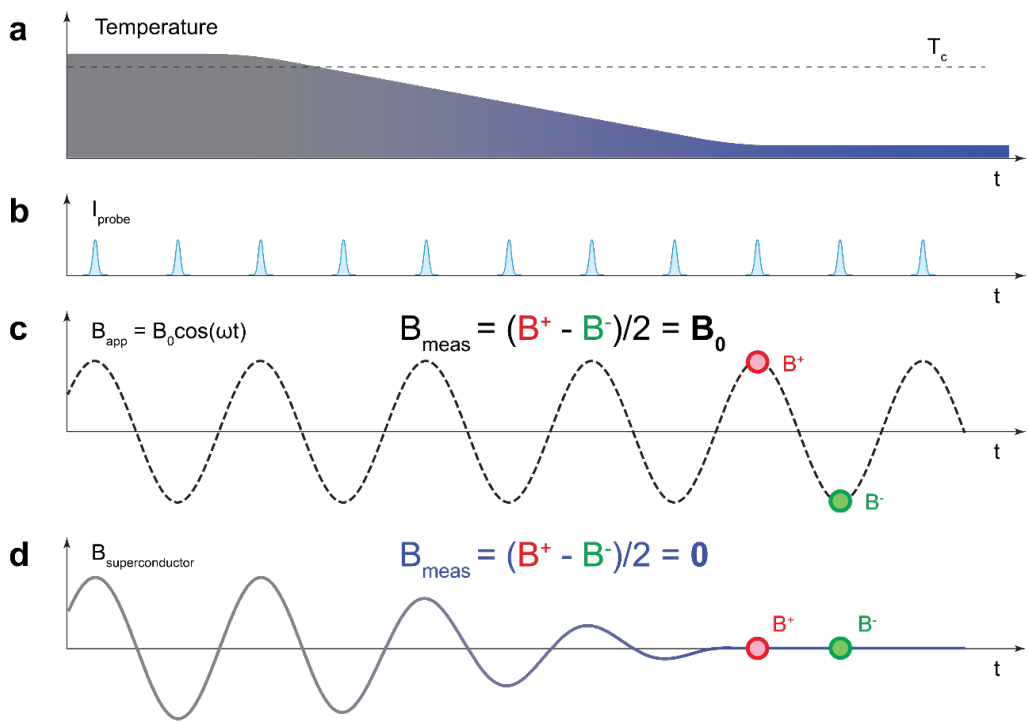

**Figure S8 | (a)** Sketch of the temporal profile of the temperature when cooling the sample below $T_c$. **(b)** Arrival time of the probe pulses sampling the magnetic field. **(c)** Externally applied magnetic field with a sinusoidal time evolution. **(d)** Magnetic shielding of the superconductor when cooling across $T_c$.

Fig. S8 qualitatively explains the equilibrium field dynamics in our experiment, which is very similar to the zero-field cooled regime. Fig. S8a shows the temperature evolution during a hypothetical experiment in which a sample is progressively cooled below $T_c$. Fig.

S8b shows the probe pulses used to sample the magnetic field at a specific point in time. When there is no sample, we expect to measure a value equal to the amplitude of the applied sinusoidal magnetic field. As described in Supplementary Information S6, the magnetic field is extracted as the difference between the polarization rotation measured when the polarity of the sinusoidal wave is positive minus the one measured when the polarity is negative. Fig. S8d schematically shows the evolution of the magnetic field upon cooling our sample across $T_c$. Initially, when the sample is above $T_c$, the magnetic field measured is almost equal to the applied magnetic field due to the higher resistivity of the sample in the normal state. As described above, the superconductor begins to partially shield the time-varying applied field after cooling across $T_c$. In this temperature regime close to $T_c$, the shielding is incomplete due to residual dissipation and the superconductor's low critical current density. Upon further cooling, the shielding improves as the superfluid density increases and the superconductor behaves like a perfect metal in which the eddy currents induced by the applied field become persistent due to the absence of dissipation. In this regime of perfect screening, the field on top of the sample is constant. Our differential measurement would yield a value of the magnetic field close to zero above the center of the superconductor. In practice, the magnetic field shielding of our pellet samples is incomplete (see the zero-field-cooled magnetization reported in Supplementary Information S9). Therefore, we measure a sizeable but partial reduction of the externally applied magnetic field on top of the superconducting patches of our sample (Fig. 2b of the Main Text).

**In Metals**

Metals' response to time-varying magnetic fields is similar to that of superconductors. The main difference is that their conductivity is finite, and electrical transport is dissipative. Therefore, the shielding currents induced in metals by time-varying magnetic fields decay with a characteristic time constant inversely proportional to their resistivity and are not persistent like in superconductors. Due to their geometry and relatively low conductivity, the decay times of eddy currents in the metallic elements in our sample holder (Supplementary Information S4) are much faster than the time scales of a few hundred Hz at which we switch the externally applied magnetic field. Therefore, the applied field changes are adiabatically slow for these currents and cause a negligible shielding response. Conversely, if we focus on the opposite limit, in which magnetic field changes happen on time scales much shorter than the decay time of eddy currents, the metal behaves similarly to a superconductor, as described in the previous subsection. That is why a Meissner response of the photo-excited state happening on ultrafast time scales would be almost entirely shielded by the metallic $K_3C_{60}$ puck not excited by the pump (Supplementary Information S11 and Fig. 3d,3e of the Main Text).

## S9. Equilibrium Magnetic Properties of $K_3C_{60}$ Pellets

This section discussed the equilibrium magnetic properties of $K_3C_{60}$ pellets. The data shown here have been measured with the experimental setup described in this manuscript and with a Quantum Design MPMS3 Superconducting Quantum Interference Device (SQuID). The two main characteristics that emerge from measurements performed on multiple samples are their poor quality compared to single crystals of $K_3C_{60}$ and a broad sample-to-sample variability.

### Sample Variability

We performed a series of SQuID magnetometry measurements at equilibrium to test the quality of the $K_3C_{60}$ powder, the possible effects of the pellet preparation procedure on the sample, and the overall reproducibility of our samples. We produced each pellet in an Ar glovebox, achieving disk-shaped samples similar to the one shown in Fig. S3.2c, which had a known diameter and thickness. While still in the glove box, we encapsulated each pellet between two 50 μm thick indium sheets pressed together to protect it from contamination during the transfer to the SQuID. The magnetic field of the SQuID was applied parallel to the disk's axis. We accounted for demagnetizing effects using the approximated expression for the demagnetizing factor of a thin cylinder with the magnetization aligned along its axis: $N \sim 1/(1+1.6/A)$, from reference [13]. In this formula, A is the ratio between the diameter and the height of the cylinder.

In Fig. S9.1, we report the intrinsic volume magnetic susceptibility extracted from data collected from six pellets produced with different batches of $K_3C_{60}$ powder. We see a marked sample-to-sample variability and notice that none of the samples measured showed a complete magnetic field expulsion ("Field Cooled") or exclusion ("Zero Field Cooled") even at the lowest temperature and magnetic field applied. For this reason, we further investigated the equilibrium properties of our $K_3C_{60}$ pellets, as described in the following subsection. The statistical quantities reported in Fig. 3c of the Main Text have been extracted from the "Field Cooled 1 mT" dataset shown in Fig. S9.1, as a proxy for the hypothetical magnetic field expulsion of the photo-excited state in the 0.5 mT externally applied field in our experiment.

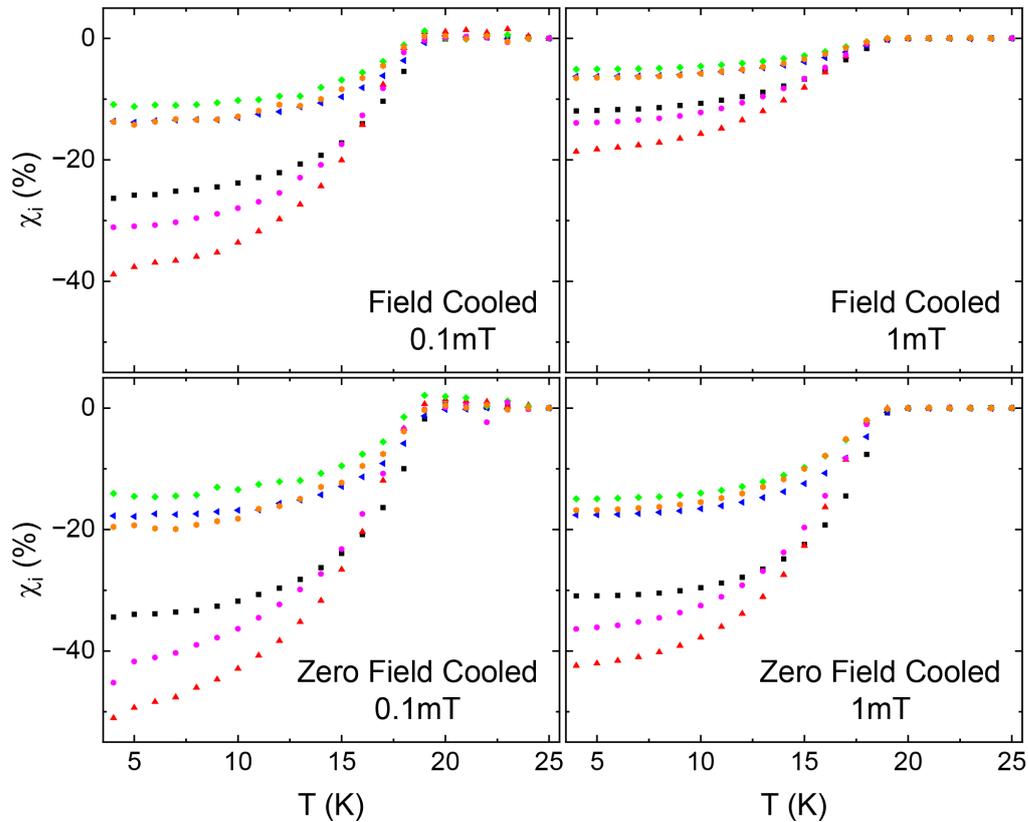

**Figure S9.1** | Intrinsic (corrected for the demagnetizing factor of the sample) magnetic susceptibility $\chi_i$ extracted from SQuID magnetization measurements performed on six different samples both in field-cooled and zero-field-cooled conditions at 0.1 mT and 1 mT externally applied magnetic field.

## Sample Granularity

We performed magnetization measurements versus the applied magnetic field to investigate further our samples' incomplete magnetic field expulsion (Fig. S9.1). We used a measurement protocol already employed to quantify the first critical field of $K_3C_{60}$ single crystals in reference [14]. The idea of this method is displayed in Fig. S9.2a-b below.

The sample is cooled from above $T_c$ in zero magnetic field to a temperature lower than $T_c$. An externally applied magnetic field is then ramped up to a specific value, x, and set to zero again. At this point, the magnetic moment trapped inside the sample is measured. Suppose the value of the externally applied magnetic field is lower than the first critical field of the sample. In that case, no magnetic flux will penetrate the sample, and no trapped magnetic moment will be measured (Fig. S9.2a).

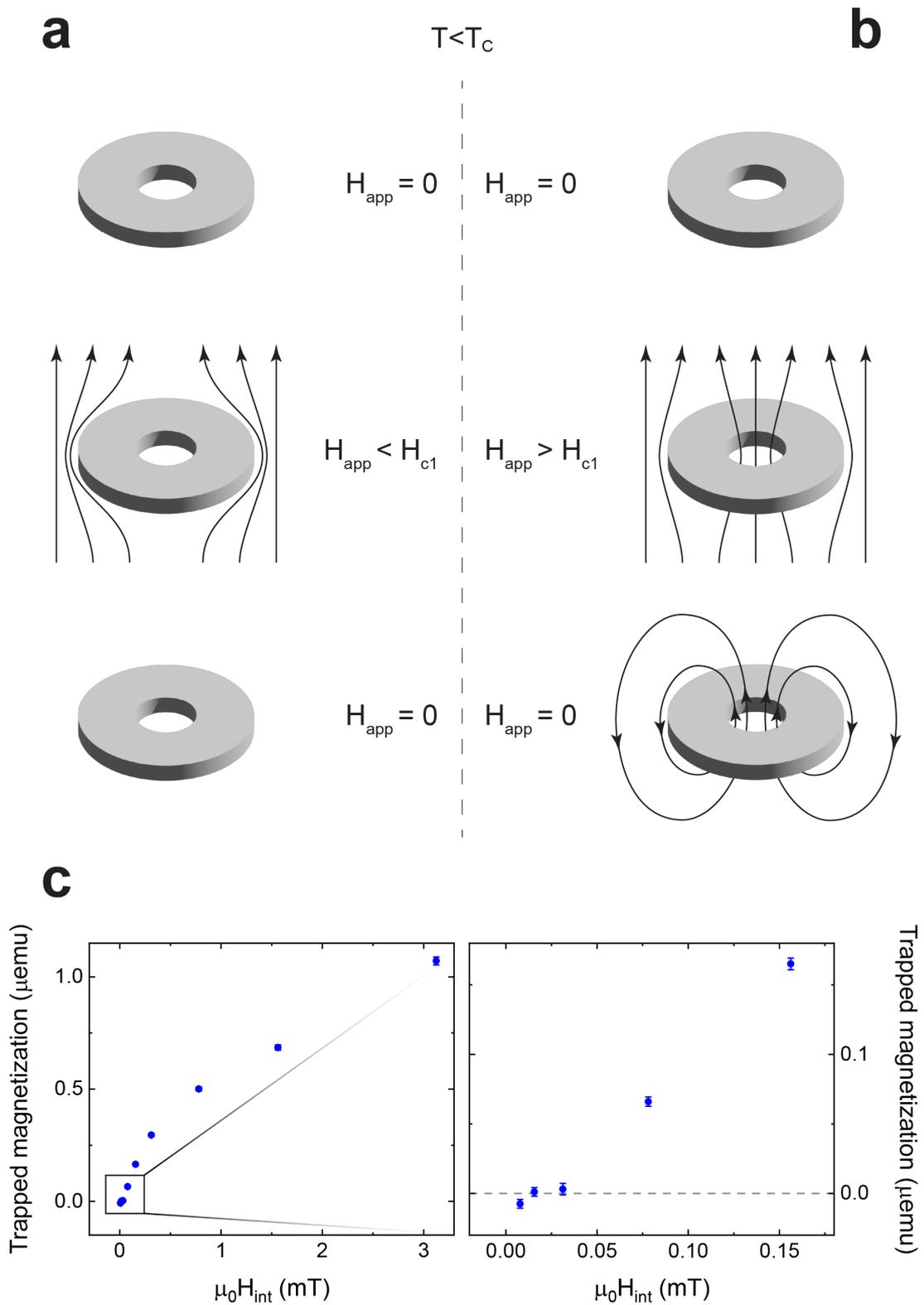

**Figure S9.2** | **(a-b)** Experimental protocol for the "trapped magnetic-moment" measurements described in reference [14]. When the externally applied magnetic field exceeds the first critical field ($H_{c1}$), the sample retains a trapped magnetic moment upon nulling again the applied field **(b)**. **(c)** Trapped magnetic moment measured with a SQuID magnetometer as a function of the applied magnetic field corrected for demagnetizing effect ($H_{int}$) at 5 K. The right panel shows a zoom-in at low magnetic fields.

Conversely, if the applied field exceeds the first critical field, some magnetic flux will penetrate the sample's volume. Upon nulling the externally applied field again, some superconducting currents will build up to keep the magnetic flux threading the sample volume constant (due to the Faraday-Lenz law), producing a trapped magnetic moment in the sample (Fig. S9.2b). The sample is then heated above $T_c$ to erase its memory, and the cycle is repeated by applying a different maximum value of the external magnetic field.

Fig. S9.2c shows the trapped magnetic moment of a $K_3C_{60}$ pellet at a temperature of 5 K for different values of the applied magnetic field corrected for the demagnetizing effects to achieve the internal field $H_{int}$. In this sample, the first critical field $\mu_0 H_{c1}$, defined above, lies between 30-80 µT.

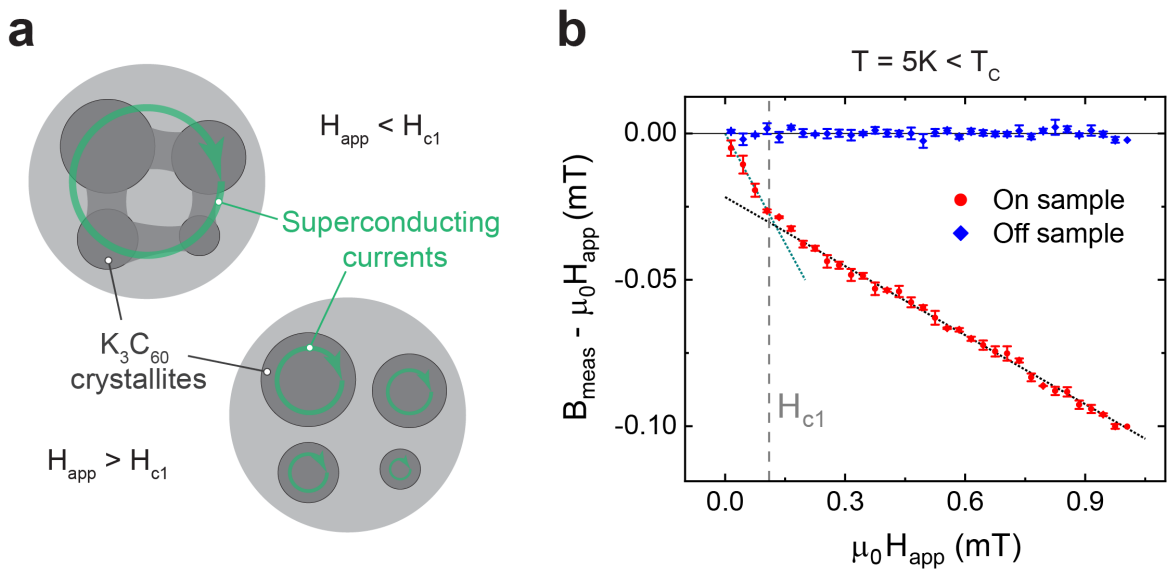

**Figure S9.3** | **(a)** Schematic picture of the sample granularity and how it affects the sample's magnetic properties. **(b)** shows equilibrium magnetization measurements of the $K_3C_{60}$ pellet displayed in Fig. 2 of the Main Text, performed with the setup described there. The sample's magnetization ($B_{meas} - \mu_0 H_{app}$) changes its slope at an amplitude equal to 120-130 µT of externally applied magnetic field, oscillating at 450 Hz. This change of slope indicates the crossing of the first critical field in the sample.

Fig. S9.3b shows the equilibrium field-dependent magnetization of the $K_3C_{60}$ pellet in Fig. 2 of the Main Text. These data have been acquired in a separate measurement using the same optical magnetometry setup (Fig. 2a). Upon increasing the externally applied magnetic field past a critical value, the slope of the magnetization changes, indicating a change in the magnetic susceptibility of the sample. This critical value lies around 130 µT, comparable to the results of the "trapped magnetic-moment" measurements discussed above, and we identify it with $H_{c1}$ of this sample. For comparison, the lower critical field extracted

by Buntar et al.[14] from similar trapped magnetic-moment measurements was 1.2 mT in monocrystalline samples, which showed a magnetic susceptibility as good as -100%.

Fig. S9.4 shows two-dimensional maps of the z-component of the local magnetic field measured at T < $T_c$ above the sample shown in Fig. 2b and with the magnetometry setup of Fig. 2a of the Main Text. The degree of spatial inhomogeneity and the magnitude of the sample's magnetic field shielding (blue) varies with the value of the externally applied magnetic field, which oscillates at 450 Hz (Supplementary Information S8). At the lowest applied field ($B_{app}$ = 0.2 mT), the measured shielding fraction ($\Delta B/B_{app}$) reached values as low as -85 %, while at the highest field ($B_{app}$ = 2 mT), it is at best -30 %. For low values of the applied magnetic field, patches appear in the sample where the magnetic shielding is significantly higher than the one measured at high values of the applied field.

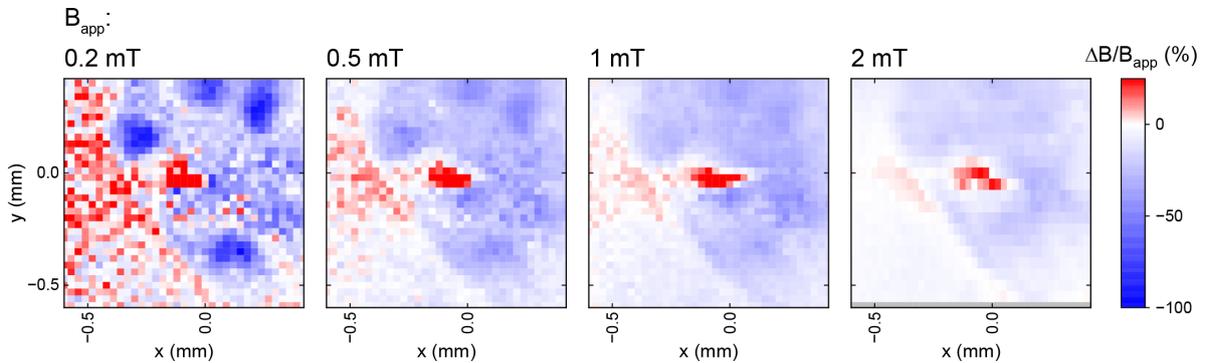

**Figure S9.4** | Two-dimensional maps of the z-component of the local magnetic field, measured as a function of the x and y position of the probe at a constant temperature T = 5 K < $T_c$ above the sample shown in Fig. 2b of the Main Text and with the same experimental setup (shown in Fig. 2a). These data have been acquired for different values of the externally applied magnetic field. An inhomogeneous reduction of the magnetic field is measured above the sample (blue), indicating the spatially inhomogeneous magnetic field shielding. The shielding fraction ($\Delta B/B_{app}$) and the shielding inhomogeneity of the sample increase as the externally applied magnetic field is reduced. At the lowest applied field ($B_{app}$ = 0.2 mT), the minimum shielding fraction measured is as low as -85 %. The red spot within the sample contour is due to an imperfection of the magneto-optic detector. The data acquired at $B_{app}$ = 0.5 mT is the same as shown in Fig. 2b of the Main Text.

These observations suggest that our powder samples are granular, with small superconducting grains connected by weak links, which break down at very low applied magnetic fields (Fig. S9.3a). Moreover, these superconducting grains seem to be connected by weak links whose strength varies spatially throughout the sample, as suggested by the formation of macroscopic islands with a lateral size of ∼ 100 μm at low values of the exter-nally applied magnetic field (Fig. S9.4). These hypotheses would explain the reduced diamagnetic susceptibility of the pellet samples and their much lower first critical field compared to the literature values for single crystals. Similar conclusions agree with the results of transport measurements performed on thin films of $K_3C_{60}$[15,16].

The strong dependence of the equilibrium properties of $K_3C_{60}$ on the externally applied magnetic field leads us to imagine that the photo-excited state also possesses a non-trivial susceptibility to the applied field. We address this question in Supplementary Information S12, showing the results of pump-probe measurements obtained at various values of the externally applied magnetic field.

## S10. Simulated Spatial Magnetic Field Distribution

The simulated distribution of the z component of the magnetic field around the whole $K_3C_{60}$ pellet at equilibrium and around the photo-excited region are shown in Fig. 2a and Fig. 3a of the Main text. They have been calculated using a magnetostatic model in COMSOL. We modeled the superconductors as disks with a uniform diamagnetic susceptibility. The magnetic field generated by the samples was calculated using COMSOL Multiphysics to solve Maxwell's equations.

An aspect ratio (diameter/thickness) of 6 has been used to simulate the equilibrium sample, which does not respect the aspect ratio of the entire puck ($\sim$ 1 mm / 30 μm $\sim$ 33). However, this value is closer to the superconducting patches measured in reality (Fig. 2b), which had a lateral size of $\sim$ 200 μm. Conversely, the pumped region was simulated as a thin disk with a diameter of 100 μm and a thickness of 400 nm, corresponding to the pump spot size and the pump penetration depth, respectively (Supplementary Information S11). The extreme aspect ratio of this geometry limits the changes to the applied magnetic field of the sample.

These simulations indicate that a sample developing a (transient) diamagnetic susceptibility would reduce the magnetic field measured with our probe on top of it.

## S11. Simulated Magnetic Dynamics

This section discusses the time-dependent finite element simulations we performed to calculate the dynamics of the possible magnetic field expulsion from the photo-excited sample in our experimental configuration (Supplementary Information S4). We used COMSOL Multiphysics to solve Maxwell's equations and evaluate the average magnetic field in a domain corresponding to the volume of the magneto-optic detector traversed by our probe pulse. This domain was 3 µm thick (as the Bi:LIGG crystal), had a diameter of 50 µm (equal to the FWHM of the probe beam), and was placed a distance of 35 µm from the excited region to account for the thickness of the $K_3C_{60}$ puck. The pumped region was simulated as a thin disk with a diameter of 100 µm (approximately equal to the FWHM of the mid-infrared pulse) and a thickness of 400 nm[10].

To faithfully simulate the dynamics of the local magnetic field around the photoexcited region, we also considered the effect of the 150 nm thick aluminum layer coating the Bi:LIGG crystal and the 30 µm thick, 1 mm diameter unperturbed $K_3C_{60}$ pellet. Due to their high conductivity, they act as a low-pass filter, reducing a possible signal generated by the sample (Supplementary Information S8). To keep the size of the simulation domain manageable, we limited the diameter of the thin aluminum film to 2 mm, slightly smaller than the actual dimension in the experiment (Fig. S4c). Note, however, that the results of the simulation are not sensitive to this parameter since it is significantly larger than the later dimension of the photo-excited region (100 µm). The conductivity of metallic $K_3C_{60}$ was set to $8·10^4$ S/m as measured in previous THz conductivity experiments. The conductivity of the 150 nm aluminum thin film was measured at DC frequency using a Van der Pauw configuration in a Quantum Design Physical Property Measurement System (PPMS). The measurement yielded a conductivity of ~$3·10^7$ S/m at 50 K (the temperature at which we performed the pump-probe experiment, see Fig. 3). We used this measured value in our simulations. We then simulated the expected magnetic field dynamics under two scenarios: (i) assuming a metastable diamagnetic response emerging after photo-excitation and (ii) a diamagnetic response persisting only for the duration of the drive. The results of these simulations are presented in Fig. 3(d,e) of the Main Text. The photo-excited area was modelled as a medium that develops a homogenous, time-dependent magnetic susceptibility $\chi_{\text{photo}}(t)$.

In the first scenario, $\chi_{\text{photo}}(t)$ is modeled as a step function with ~1 ps fall time reaching final values of the diamagnetic susceptibility, $\chi_{\text{photo}}(t \to +\infty)$, equal to the average equilibrium field-cooled magnetic susceptibility of $K_3C_{60}$ at 4 K and 16 K, see Fig. 3c of the Main Text and Supplementary Information S9. The resulting time traces of $\chi_{\text{photo}}(t)$ are displayed as the blue and red lines in Fig. 3d, top panel, representing the 4 K and 16 K cases, respectively. We performed additional simulations to account for the sample variability assigning to $\chi_{\text{photo}}(t \to +\infty)$ the maximum and minimum values of the equilibrium diamagnetic susceptibility measured at 4 K and 16 K (Supplementary Information S9). These values correspond to the limits of the shaded areas in Fig. 3d, top panel.

The corresponding magnetic field changes are shown in the lower panel of Fig. 3d of the Main Text. Regardless of the final value $\chi_{\text{photo}}(t \to +\infty)$, the traces show a long-lived magnetic field change appearing after an initial drop with a time constant of ~120 ps. As discussed above, this decay time is due to the inductive response of the unperturbed part of the metallic K$_3$C$_{60}$ pellet and the aluminum thin film deposited on the Bi:LIGG detector (see Supplementary Information S4). The values of long-lived magnetic field change ($\Delta B$) in the different cases discussed above are summarized in Table S11:

| $T = 4\ K$ | | $T = 16\ K$ | |
|---|---|---|---|
| $\chi_{\text{photo}}(t \to +\infty)$ | $\Delta B\ (nT)$ | $\chi_{\text{photo}}(t \to +\infty)$ | $\Delta B\ (nT)$ |
| $\chi_{\text{i, max}} = -0.05$ | $-50$ | $\chi_{\text{i, max}} = -0.02$ | $-20$ |
| $\chi_{\text{i, ave}} = -0.105$ | $-115$ | $\chi_{\text{i, ave}} = -0.04$ | $-40$ |
| $\chi_{\text{i, min}} = -0.185$ | $-230$ | $\chi_{\text{i, min}} = -0.055$ | $-60$ |

**Table S11** | Calculated long lived magnetic field change $\Delta B$ following a metastable quench of the photo-induced diamagnetic susceptibility $\chi_{\text{photo}}(t)$ to values equal to the maximum, average and minimum measured equilibrium magnetic susceptibility $\chi_i$ at 4 K and 16 K (Supplementary Information S9).

In the second scenario $\chi_{\text{photo}}(t)$ follows a Gaussian pulse peaking at $\chi_{\text{photo}}(t = 0) = \chi_{i,ave}(4\ K)$ (solid blue line in Fig. 3e, top panel). Due to the inductive response of nearby metallic elements, the maximum amplitude of the magnetic field change is severely reduced and is ~-2 nT, well below our experimental resolution (solid blue line Fig. 3e, lower panel). The limits of the the shaded blue band in Fig. 3e assume $\chi_{\text{photo}}(t \to +\infty)$ to equal the maximum and minimum measured diamagnetic susceptibilities at equilibrium at 4 K. Finally, we assessed the impact on the temporal resolution of the experiment of the thin aluminum coating.

The relative contributions to the overall magnetic field shielding of the equilibrium K$_3$C$_{60}$ sample and aluminum thin film is shown in Fig. S11. This figure shows the simulated results performed under the assumption of a metastable magnetic response of the photo-excited state (see below). The black line represents the case in which the effect of both the K$_3$C$_{60}$ puck and aluminum film is simulated. The blue line represents the case in which only the K$_3$C$_{60}$ pellet is accounted for in the model. By fitting the curves with exponentially decaying functions (dashed lines in the same figure), we can extract a time constant of ~120 ps and ~40 ps in these two cases discussed. With the aluminum film, the time resolution is worsened by a factor of three, which does not qualitatively impact the simulated magnetic dynamics discussed above, but the noise of the probe is significantly reduced.

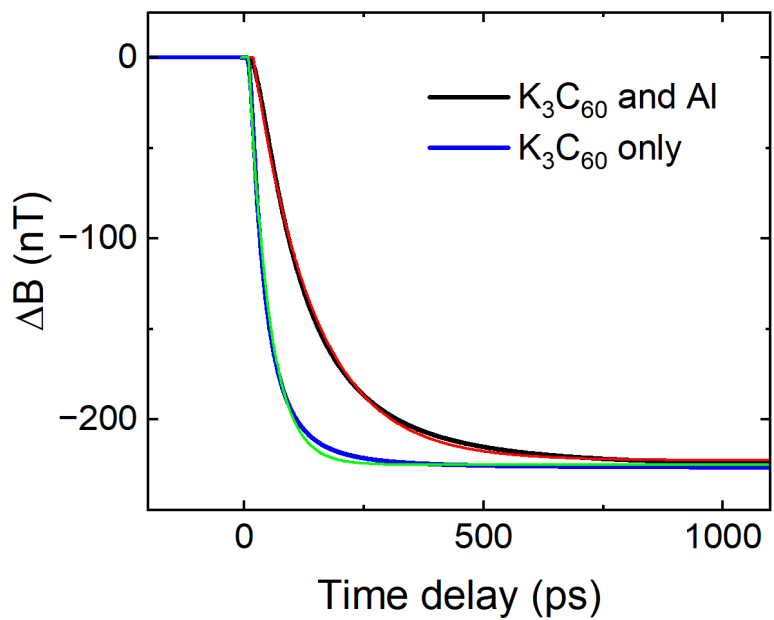

**Figure S11** | Simulated magnetic field dynamics for the case of a metastable susceptibility quench with amplitude -0.185 when only the metallic non-excited $K_3C_{60}$ puck is considered (blue solid line) and when both the puck and the aluminum film are considered (black solid line). The red and green lines are single time-constant exponential decay functions used to fit the data.

## S12. Magnetic Field Dependence of Pump-probe Measurements

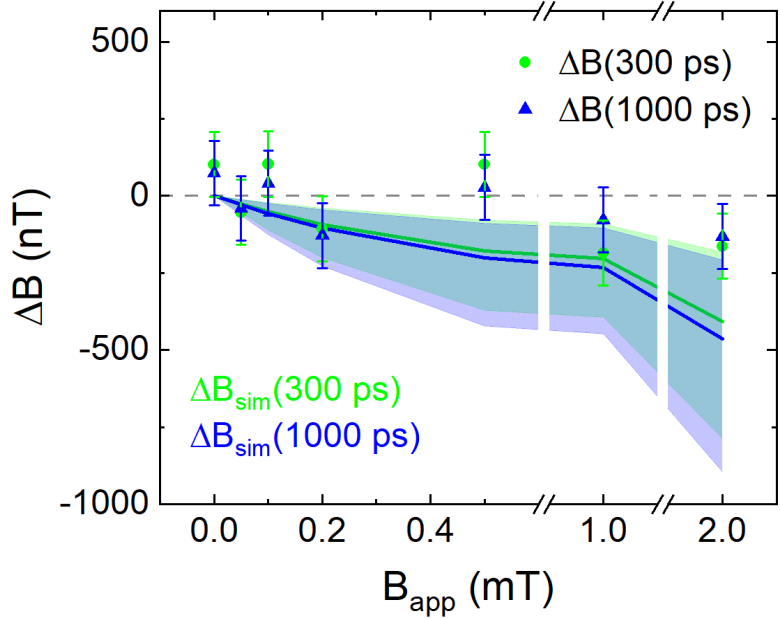

**Figure S12** | Applied magnetic field dependence of the pump-induced changes in the z-component of the local magnetic field ΔB measured at selected pump-probe time delays. The solid lines and shaded areas show the simulation results of the same quantity (ΔB$_{sim}$). The error bars denote the standard error of the mean.

Fig. S12 presents the results of pump-probe measurements performed at different values of the externally applied magnetic field with a pump fluence of 20 mJ/cm². The data points in the figure show the pump-induced magnetic field change measured above the K$_3$C$_{60}$ at seven different values of the externally applied magnetic field for two selected pump-probe delays, of 300 ps and 1000 ps. The solid lines and the shaded bands are the magnetic field change resulting from simulations assuming a metastable photo-induced diamagnetic susceptibility. The values used for the pump-induced long-lived diamagnetic susceptibility are summarized in Table S12:

| $\chi_{\text{photo}}(t \to +\infty)$ | $B_{app} \leq 0.1mT$ | $B_{app} = 0.2mT$ | $B_{app} = 0.5mT$ | $B_{app} \geq 0.5mT$ |
|---|---|---|---|---|
| $\chi_{\text{i, max}}$ | −0.11 | -0.103 | -0.083 | −0.05 |
| $\chi_{\text{i, ave}}$ | −0.225 | -0.21 | -0.17 | −0.105 |
| $\chi_{\text{i, min}}$ | −0.39 | -0.367 | -0.3 | −0.185 |

**Table S12** | Values of the long-lived diamagnetic susceptibility, $\chi_{\text{photo}}(t \to +\infty)$, used at different applied magnetic fields in Fig. S12.

Similarly to what was done in Fig. 3d of the Main Text and as described in Supplementary Information S11, we simulated the magnetic signal after photo-excitation, assuming a metastable pump-induced diamagnetic susceptibility, $\chi_{\text{photo}}(t)$, in the pumped volume. We assigned to the long-lasting diamagnetic susceptibility, $\chi_{\text{photo}}(t \to +\infty)$, values extracted from the equilibrium field cooled diamagnetic susceptibilities measured at $T = 4\,K \ll T_c$ and at $B_{\text{app}} = 0.1, 1\,mT$ (see Fig. S9.1). These values are summarized in Table S12. For $B_{\text{app}} \leq 0.1\,mT \sim H_{c1}$, we used the maximum, minimum, and average (upper edge, lower edge of the shaded bands, and solid line in Fig. S12, respectively) values of the 4 K field-cooled diamagnetic susceptibility measured at $0.1\,mT$. For $B_{\text{app}} \geq 1\,mT$, we used analogous data measured at $1\,mT$. Finally, we extracted the values of $\chi_{\text{photo}}(t \to +\infty)$ at $B_{app} = 0.2, 0.5\,mT$ by linearly interpolating the equilibrium diamagnetic susceptibilities measured at 0.1 and $1\,mT$. For a better comparison with the measured data points at 300 and 1000 ps pump-probe delay, the simulated magnetic field change in Fig. S12 is also reported at 300 (green line and shaded area) and 1000 ps (blue line and shaded area).

In the low field range ($B_{app} \leq 0.2\,mT$), the data do not provide a conclusive answer, given the value of the error bars in Fig. S12. Conversely, at higher values of the applied magnetic field, the data confirm the results shown in Fig. 3b of the Main Text, that no pump-induced signal compatible with an equilibrium 4 K-like diamagnetic response of the photo-excited state can be detected.

# References (Supplementary Information)